\documentclass[manuscript,screen,review=false]{acmart}

\AtBeginDocument{%
  \providecommand\BibTeX{{%
    \normalfont B\kern-0.5em{\scshape i\kern-0.25em b}\kern-0.8em\TeX}}}

\setcopyright{acmcopyright}
\copyrightyear{2021}
\acmYear{2021}
\acmDOI{XX.XXXX/XXXXXXX.XXXXXXX}

\acmJournal{CSUR}
\acmVolume{30}
\acmNumber{3}
\acmArticle{102}
\acmMonth{3}

\begin{document}

\title{A Systematic Review of Security in the LoRaWAN Network Protocol}

\author{Poliana de Moraes}
\email{poliana.moraes@embraer.com.br}
\orcid{0000-0001-5429-7936}
\affiliation{%
  \institution{Federal University of São Paulo}
  \city{São José dos Campos}
  \state{São Paulo}
  \country{Brazil}
}

\author{Arlindo Flavio da Conceição}
\email{arlindo.conceicao@unifesp.br}
\affiliation{%
  \institution{Federal University of São Paulo}
  \city{São José dos Campos}
  \state{São Paulo}
  \country{Brazil}
}


\begin{abstract}
  The age of the Internet of Things is adding an increasing number of new devices to the Internet and is expected to have fifty billion connected units by 2021. These form an extensive network that may have multiple points where there is a risk of attacks that can compromise the entire system. This paper has conducted a systematic review of security in LoRaWAN protocol specification versions 1.0 and 1.1 by locating its vulnerabilities and determining what measures can be taken for improvement and how they can be checked or tested. The review identifies nineteen areas of vulnerability in the LoRaWAN protocol and shows that the current studies focus on specification version 1.0, key management, and authentication procedures.
\end{abstract}

\ccsdesc[500]{Security and privacy~Security protocols}

\keywords{systematic review, Internet of Things, LoRaWAN}

\maketitle

\section{Introduction}
The age of the Internet of Things is adding an increasing number of new devices to the Internet and is expected to have 50 billion connected units by 2021. In this context, LoRaWAN protocol is a wireless wide-area network in which the architecture and operation implement system-defined properties. The first LoRaWAN specification was released in October 2015, and the second and most recent one was released in October 2017. It is characterized by low operational costs, optimal power consumption, a high number of connected devices, and long-range communication. LoRaWAN is an attractive option for Internet of Things applications because of the following features: open standards, the provision of off-the-shelf and low-cost platforms, operating services in unlicensed industrial, scientific and medical frequencies, and a private network alternative \cite{comparative_LPWAN}. 

Since the LoRaWAN protocol is an Internet of Things application, it forms part of an extensive interconnected system that has multiple points that are vulnerable to attacks. These attacks aim to access, block, modify, and corrupt data or services. They can cause significant disruption to businesses and companies that do not have suitable security mechanisms in place. In addition, successful attacks can harm the reputation of a company, as well as cause a loss of sensitive data or a violation of intellectual property. Because of this, security is currently considered a crucial issue. The objective of this paper is to conduct a systematic review of security in LoRaWAN protocol specification versions 1.0 and 1.1 by locating its vulnerabilities and determining what measures can be taken for its improvement and how they can be checked or tested. This paper is divided into several topics, which can be described as follows: in Topic 2, there is an examination of the LoRaWAN protocol design and security mechanisms, while Topic 3 describes the systematic review methodology employed in this paper. The protocol for the systematic review is defined in Topic 4. The selection of studies for research is investigated in Topic 5 under the requirements of the systematic review protocol. In Topic 6, the extracted data are compiled based on the defined questions. In Topic 7, a summary is provided, which includes key points, significant results and recommendations for future work in the field.

\section{LoRaWAN}
LoRa Alliance is an open, nonprofit organization founded in 2015 by leading technological industrialists. Its aim is to standardize low-power wide-area networks to implement the Internet of Things, machine to machine, and industrial application technologies. As shown in Figure, the LoRaWAN protocol stack consists of three layers: the physical layer, the data link layer, and the application layer. The application layer deals with the data that are transmitted through the network. The data link layer implements the LoRaWAN protocol that is defined by the LoRa Alliance. The physical layer implements the LoRa radio frequency technology from Semtech \cite{lorawan1.1}. 

\begin{figure}[h]
  \centering
  \includegraphics[width=\linewidth]{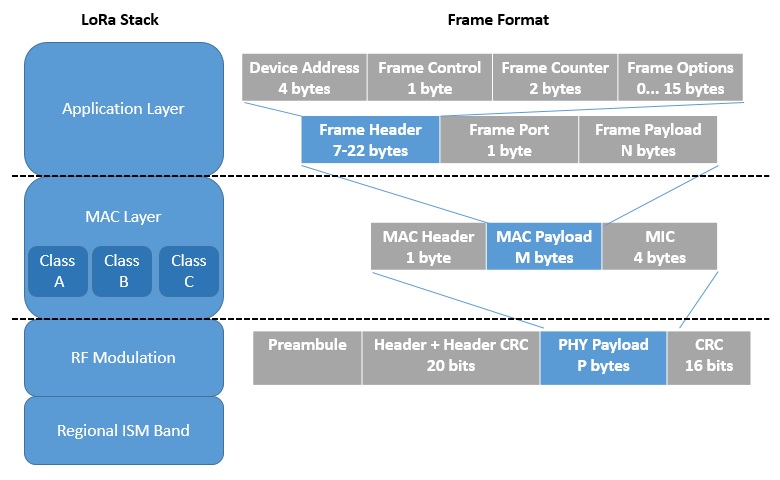}
  \caption{Stack and Frame Format}
  \Description{The relationship between the LoRaWAN stack and frame format.}
\end{figure}

The LoRaWAN protocol specifies a star network topology and ensures that each component has its own particular role in the system. An extensive number of end devices can form it. End devices are either fixed or mobile entities that are responsible for collecting and transmitting data via LoRa radio frequency to the gateway. As there is no specific gateway combined with end devices, it can transmit the data to one or more gateways. The gateway is responsible for forwarding messages, via standard IP, and making connections from the end devices to the related network server. The network server is responsible for managing the entire network. It analyzes the payload and checks and eliminates the redundant messages. It also checks the frame authenticity and counters. Then, it routes the message to the defined application server. The application server is responsible for processing the data and taking any necessary action; it is the end server. The entire architecture is shown in Figure.

\begin{figure}[h]
  \centering
  \includegraphics[width=\linewidth]{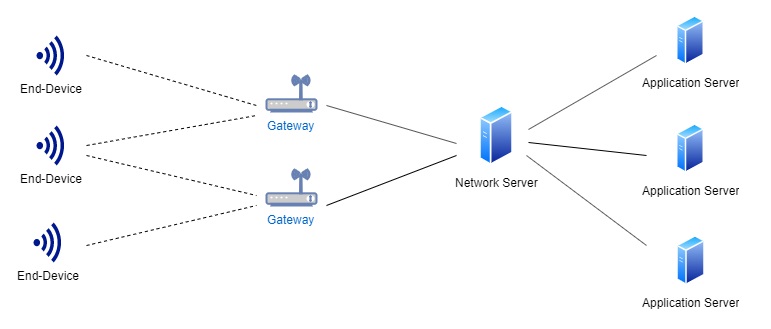}
  \caption{LoRaWAN Architecture}
  \Description{Components of the LoRaWAN network architecture specification version 1.1.}
\end{figure}

End devices can be divided into three categories and must at least fulfill the functionality of Class A. 
Class A end devices are battery-powered sensors, and downlink communication occurs in two short windows after an uplink window; this communication strategy leads to optimal power consumption. Class B end devices are battery-powered actuators, which in addition to the two short downlink windows available in class A, also make it possible to schedule a specific time window for downlink communication. Class C end devices are the main powered actuators. They have a constantly open downlink window that only closes when the end device is transmitting data, even though high-power consumption is expected. Uplink messages are routed from the end device to the server, and downlink messages are routed from the server to the end device. Uplink communication is expected to be the predominant direction owing to power consumption restrictions in the Internet of Things applications. In the case of uplink communication, the end device establishes a connection with the gateway and only sends data when it is necessary, as it directly affects the end device battery lifetime \cite{lorawan1.1}.

LoRaWAN provides data encryption from the end devices to the application server. The basic properties that are supported by LoRaWAN security are mutual authentication, integrity protection, and confidentiality. Mutual authentication is assigned to guarantee the identity of the end devices. This ensures that only authorized end devices can join the network. Integrity protection is provided to guarantee that the messages have not been altered; this is carried out by checking the validity of the messages. Confidentiality is required to ensure that third parties will not read the messages.

LoRaWAN relies on AES cryptography algorithms combined with operational mode. The application payload is encrypted in the application layer by 128-bit symmetric keys and uses a frame counter to prevent message replay. Moreover, a message integrity code is created with 128-bit symmetric keys in the data link layer. The security keys specified in LoRaWAN version 1.1 are shown in Figure. It should be noted that the protocol employs two layers of security, one for the network and one for the application. Network encryption ensures the authenticity of the end device in the network, while the application payload encryption ensures authenticity and integrity.

\begin{figure}[h]
  \centering
  \includegraphics[width=\linewidth]{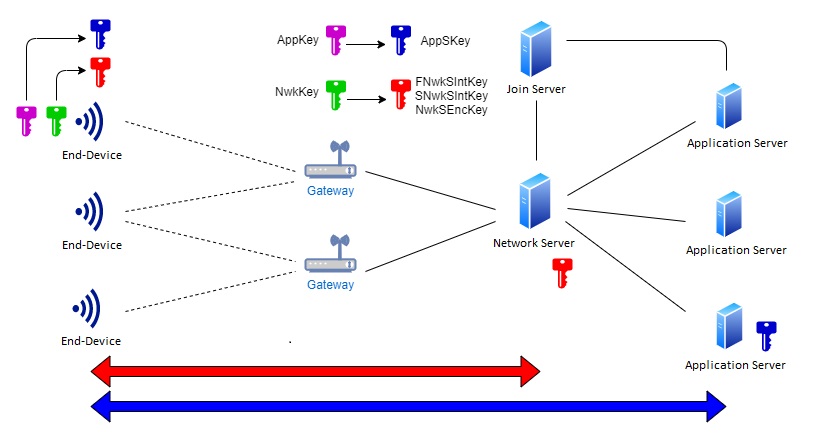}
  \caption{Security Keys}
  \Description{The root and session keys of the LoRaWAN specification version 1.1.}
\end{figure}

There are two procedures for carrying out end device activation in a LoRaWAN network: activation by personalization and over-the-air activation. Activation by personalization does not follow the join procedure; the session keys and related parameters are provisioned in the end device. Over-the-air activation dynamically generates session keys from root keys during the join procedure. There are some differences between specification versions 1.0 and 1.1 in these processes.

In specification version 1.0, the AppKey root key is recorded in the end device during the manufacturing or commissioning process. When the end device is connected to the network for the first time, it sends a join request message to the network server. The network server checks the authenticity of the end device and uses the root keys and related parameters to derive the session keys and the end device address. Then, it sends the join accept message with the end device address and a dataset that the end device needs to derive the session keys, which are AppSKey and NwkSKey. Finally, a session can be initiated, and data exchanged, as shown in Figure \cite{lorawan1.0}.

\begin{figure}[h]
  \centering
  \includegraphics[width=\linewidth]{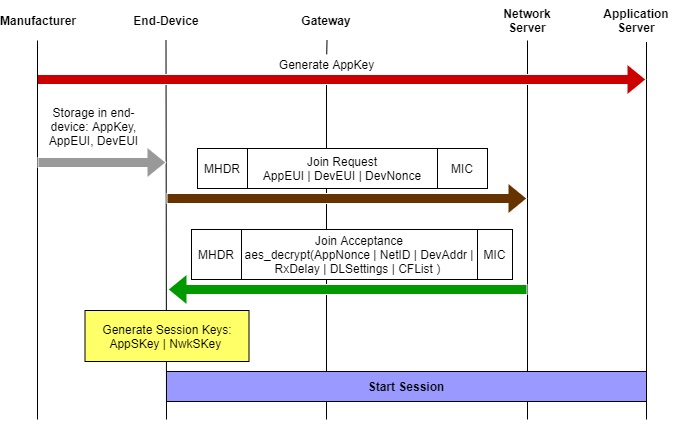}
  \caption{Over-The-Air Activation Version 1.0}
  \Description{The procedure of over-the-air activation according to LoRaWAN specification version 1.0.}
\end{figure}

In specification version 1.1, AppKey and NwkKey root keys are recorded in the end device during the manufacturing or commissioning process. When the end device is connected to the network for the first time, it sends a join request message to the network server that will then forward it to the linked join server. The join server is a new entity introduced in specification version 1.1 and is responsible for handling the end device authentication. The join server checks the end device authenticity and uses the root keys and related parameters to derive the session keys and the end device address. Then, the join server sends the join accept message with the end device address and a dataset that the end device needs to derive the session keys, which are AppSKey, FNwkSIntKey, SNwkSIntKey, and NwkSEncKey. Finally, a session can be initiated, and data are exchanged, as shown in Figure \cite{lorawan1.1}.

\begin{figure}[h]
  \centering
  \includegraphics[width=\linewidth]{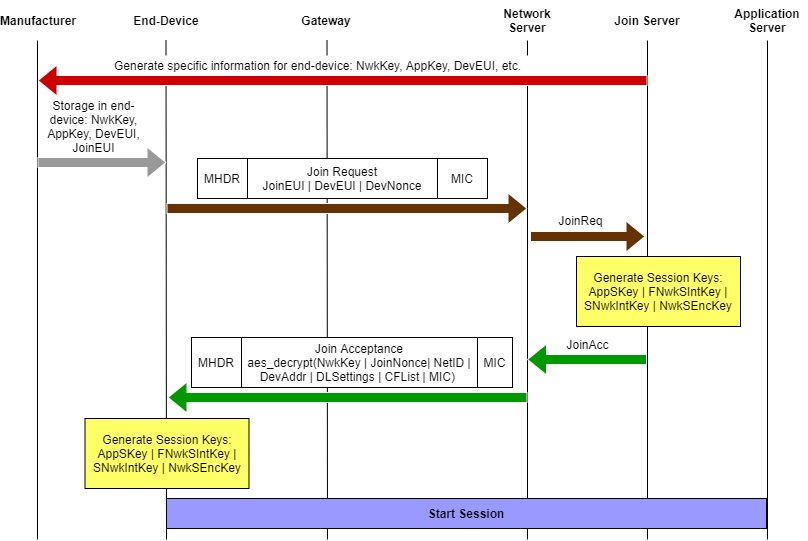}
  \caption{Over-The-Air Activation Version 1.1}
  \Description{The procedure of over-the-air activation according to LoRaWAN specification version 1.1.}
\end{figure}

\section{Methodology}

A systematic review is a methodology that is employed to appraise the importance of currently available research in a particular area. The following analysis was conducted per the guidelines drawn up by Kitchenham. The procedure is divided into 3 phases: planning, conducting, and reporting \cite{kitchenham}.

In the planning phase, an attempt is made to ascertain if the systematic review is needed or whether there is any other equivalent or similar publication that already provides the expected results. Then, the questions, strings, and review protocol are defined, which involves describing why and how the review will be carried out. In the conducting phase, the study is carried out following the protocol defined in the planning phase. In the first step, research is undertaken with the aid of string and search parameters. Then, an evaluation is made to determine which study can assist in answering the questions and fit the established criteria. Finally, an in-depth analysis is carried out for each study to extract the data. In the reporting phase, the extracted data are compiled to generate relevant information that can be used to answer the prepared questions. The entire process is shown in Figure.

\begin{figure}[h]
  \centering
  \includegraphics[width=\linewidth]{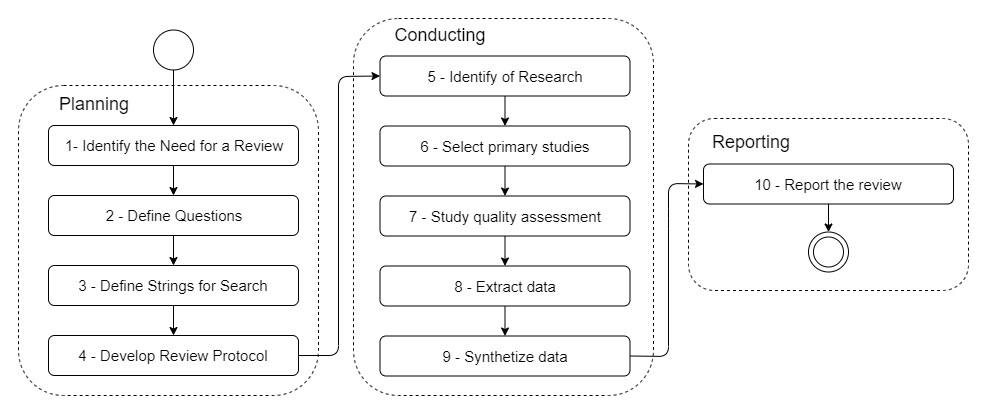}
  \caption{Systematic Review Flowchart}
  \Description{Workflow of the activities to perform a systematic review.}
\end{figure}

\section{Planning Phase}
\subsection{Establishing if there is a Need for a Review}

A search of the following string was performed in Google Scholar digital library on January 6, 2020, to find out if any systematic review has been already published about security in the LoRaWAN protocol: \textit{(``Security'' OR ``Cybersecurity'') AND (``LoRa'' OR ``LoRaWAN'') AND (``Systematic Review'' OR ``Review'' OR ``SLR'' OR ``Survey'')}. Since no systematic review related explicitly to LoRaWAN security was found in the results of this search, this systematic review was regarded as necessary.

\subsection{Defining Questions}

In this stage, the questions that will be answered about the LoRaWAN protocol are defined. The questions that are included in this systematic review are as follows:
\begin{itemize}
\item Q1 - What are the security vulnerabilities of LoRaWAN? \\
Objective: To identify the list of requirements for ensuring security vulnerability in LoRaWAN.

\item Q2 - Is there any scheme for improving security in LoRaWAN? \\
Objective: To seek mechanisms or solutions that can improve LoRaWAN security based on the requirements specified in Q1.

\item Q3 - Which of LoRaWAN's security mechanisms and vulnerabilities have been checked or tested? \\
Objective: To find out which tests can determine the kinds of vulnerability and how the plans for improvement should be carried out.

\item Q4 - For which specification version the requirements are designed? \\
Objective: To find out if the requirements are related to version 1.0, released in October 2015, or version 1.1, released in October 2017. 
\end{itemize}

\subsection{Define Strings for Search}

The keywords are extracted from the review questions and are \textit{Cybersecurity OR Security, LoRa OR LoRaWAN, Sensor Network}. Different strings were generated from the keywords to evaluate their degree of efficiency in finding significant studies, and a search was carried out in the Google Scholar digital library. Then, the first ten results were assessed and classified in terms of whether they were relevant or not, as shown in Figure. The following strings were evaluated, and the results are as follows:
\begin{itemize}
  \item \textit{Security AND (LoRa OR LoRaWAN)} returned 6 relevant studies
  \item \textit{(Security OR Cybersecurity) AND (LoRa OR LoRaWAN)} returned 6 relevant studies
  \item \textit{Security AND (LoRa OR LoRaWAN) AND Network} returned 6 relevant studies
  \item \textit{(Security OR Cybersecurity) AND (LoRa OR LoRaWAN) AND Network} returned 7 relevant studies
  \item \textit{(Security OR Cybersecurity) AND (LoRa OR LoRaWAN) AND Sensor} returned 5 relevant studies
  \item \textit{(Security OR Cybersecurity) AND (LoRa OR LoRaWAN) AND Sensor Network} returned 3 relevant studies
  \item \textit{(Security OR Cybersecurity) AND (LoRa OR LoRaWAN) AND (Sensor OR Network)} returned 6 relevant studies
\end{itemize}

\begin{figure}[h]
  \centering
  \includegraphics[width=\linewidth]{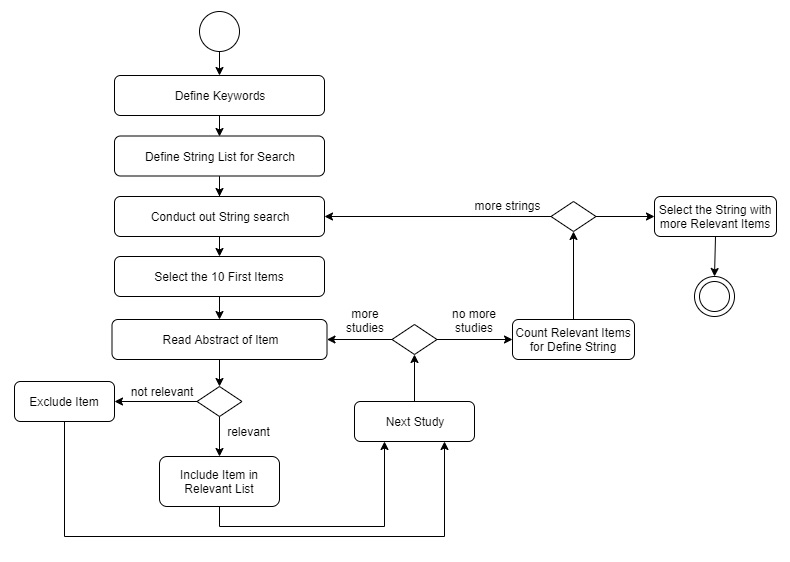}
  \caption{String Definition Flowchart}
  \Description{Workflow of the activities to perform string definitions for a systematic review.}
\end{figure}

The strings that returned the most relevant results were selected to carry out this systematic review and are as follows: \textit{(Security OR Cybersecurity) AND (LoRa OR LoRaWAN) AND Network}.

\subsection{Preparation of the Review Protocol}

This review aims to determine a) the security vulnerabilities of LoRaWAN, b) the plans to improve LoRaWAN security resilience, and c) how these factors can be checked or tested. The search is limited to the scientific articles published in journals, transactions, and scientific events so that optimal results can be obtained from this review. The selected resources are recognized in digital libraries with peer review policies and procedures, which makes it possible to ensure the quality of the input data. These are Springer Link, Science Direct, IEEE Xplore, and ACM. The search is conducted in their web service for search. The selected language is English, as the most important publications on this topic are written in this language. Since the LoRa Alliance, which is the organization responsible for the standardization of LoRaWAN, was founded in June 2015, a filter by the current year is formed to search among the publications since 2015.

The selection criteria for determining which study should include an item for detailed analysis are as follows: 
\begin{enumerate}
  \item The study must be related to the LoRaWAN protocol.
  \item The study must discuss security.
  \item The study must conduct technical analysis.
\end{enumerate}
   
The criteria for excluding an item for detailed analysis are as follows:
\begin{enumerate}
  \item The study is not related to LoRaWAN.
  \item The study is not related to security.
  \item The study only describes security in LoRaWAN.
  \item The study only cites vulnerabilities from other related works.
  \item The study only employs LoRaWAN as a security protocol.
  \item The study or item depends on physical access.
  \item The study or item is only related to the radio frequency signal in the physical layer.
\end{enumerate}

\section{Conducting Phase}
\subsection{Defining the Research}

The search was carried out following the guidelines of the protocol on January 6, 2020. In Springer Link, the search took place on \url{https://link.springer.com}. The string was inserted in the main search bar. A filter was set up for the years from 2015 and another one for the publication, which included a selected article and conference paper. It returned seven studies. In Science Direct, the search was carried out on \url{https://www.sciencedirect.com}. The string was inserted in the title, abstract, and author's specific-keyword search bar. A filter was set up for the years from 2015 and another for the publication, which included a selected research article and review article. It returned six studies. In ACM digital library, the search was done on \url{https://dl.acm.org}. The string was inserted in the search bar with the filters for the abstract and the year of publication from 2015 to 2020. It returned twenty-two studies. In IEEE digital library, the search was carried out on \url{https://ieeexplore.ieee.org}. The string was inserted in the main search bar. A filter was set up for the years from 2015 to 2020 and another one for the publication, consisting of the selected conferences, magazines, and journals. It returned eighty-eight studies. In total, one hundred and twenty-three studies had to be evaluated in this systematic review.

\subsection{Select Primary Studies}

In this step, the title, abstract, and keywords of all the studies were read and evaluated. Then, the primary studies were selected based on the inclusion and exclusion criteria. As a result, seven studies were excluded because they were duplicated. Twenty-seven studies were excluded because they were not related to the LoRaWAN protocol. Forty-five studies were excluded because they were unrelated to security, and 8 studies were excluded because they were not associated with LoRaWAN and security. Thus, a total of 87 studies were excluded in this phase. It can be concluded that there are thirty-nine primary studies, namely, two from Springer Link, three from Science Direct, five from ACM digital library, and twenty-nine from IEEE.

\subsection{Data Extraction}

In this stage, the primary studies were read in their entirety, and the data extraction was carried out concerning the parameters that define the method of answering the identified questions.

During the data extraction, eight studies were excluded. Oniga {\it et al.} \cite{36_secure_lorawan_sensor_networks} implemented a secure network architecture to which LoRaWAN is applied and where security features comprise a small section of the studies; however, it does not clearly define the LoRaWAN vulnerabilities. Luki{\'c} {\it et al.} described the security features of the LoRaWAN specification without conducting a security analysis \cite{53_dataflow_IoT_application} . Winderickx {\it et al.} \cite{65_endtoend_secured_comm}, Ozyilmaz and Yurdakul \cite{66_blockchain_swarm_LoRa}, Webb and Hume \cite{69_NIST_IoT}, and Dmitrievich and Dmitrievich \cite{81_Integration_LPWAN} applied LoRaWAN as an end-to-end security protocol in their solutions but without analyzing LoRaWAN security. Behrad {\it et al.} discussed access control for the Internet of Things application and also analyzed LoRaWAN but did not provide an evaluation of vulnerability \cite{78_network_access_control}. Chen {\it  {\it et al.} } created a new encryption algorithm, but their focus was on the processing restrictions of the LoRaWAN network and not on the security \cite{121_sessin_key_generation}. The final result was that thirty-one primary studies were used for data extraction, namely, two from Springer Link, three from Science Direct, five from ACM digital library, and twenty-one from IEEE.

\subsection{Study Quality Assessment}

A quality assessment was carried out to determine the value of the extracted data related to the defined questions.
\begin{enumerate}
  \item Does the study focus on LoRaWAN security? \\
  $87.1$\% are positive, $12.9$\% are moderately positive, and $0$\% are negative.
   
  \item Does the detailed study make any improvement to LoRaWAN security? \\
  $86.4$\% are positive, $13.6$\% are moderately positive, and $0$\% are negative.
   
  \item Does the study examine the test tools and procedures in detail? \\
  $83.3$\% are positive, $16.7$\% are moderately positive, and 0\% are negative.
\end{enumerate}

\subsection{Data Synthesis}

The data extraction is carried out under the following categories:
\begin{itemize}
\item Identification of study including title, year of publication, authors and digital library;
\item Detected security vulnerabilities;
\item LoRaWAN specification version;
\item Planned improvement about the related vulnerability; and 
\item Checking or testing was carried out, including the question of whether it is related to vulnerability or recommended improvements.
\end{itemize}

\section{Reporting Phase}

In this phase of the systematic review, the extracted data were compiled and evaluated.  One topic was assigned for each vulnerability when at least one answer was found for the defined questions.

\subsection{A lack of end-to-end integrity protection}

\begin{itemize}
\item Q1 - Eldefrawy {\it et al.} \cite{01_formal_security_analysis_LoRaWAN}, D\"onmez and Nigussie \cite{03_security_LoRaWANv1.1_scenarios}, Gladisch {\it et al.} \cite{24_securely_IoT_lorawan}, Yang {\it et al.} \cite{37_security_vulnerabilities_lorawan}, Skorpil {\it et al.} \cite{59_IoT_security_overview_practical_demonstration} and Naoui {\it et al.} \cite{98_lorawan_framework_smarthome} found that checks on the message integrity code is calculated and carried out at the network server, which means that data integrity is unprotected from the network server to application server, leaving the data vulnerable to integrity attacks.

\item Q2 - Gladisch {\it et al.} proposed applying standardized network security solutions, such as SSL/TLS, between the network and application servers \cite{24_securely_IoT_lorawan}. Yang {\it et al.} \cite{37_security_vulnerabilities_lorawan} and Naoui {\it et al.} \cite{98_lorawan_framework_smarthome} proposed running the integrity check utility at the application server or including another integrity check in the protocol to add end-to-end integrity protection.

\item Q3 - Yang {\it et al.} launched a man-in-the-middle attack between a network server and an application server, which led to several changes in the message such as data value, end device address, and counters \cite{37_security_vulnerabilities_lorawan}.

\item Q4 - LoRaWAN specification versions 1.0 and 1.1.

\end{itemize}

\subsection{Join and rejoin request messages are not encrypted}
\begin{itemize}
\item Q1 - Eldefrawy {\it et al.} \cite{01_formal_security_analysis_LoRaWAN}, Naoui {\it et al.} \cite{35_third_party_key_management}, Na {\it et al.} \cite{44_countermeasure_replay_attack}, Benkahla {\it et al.} \cite{49_analysis_lorawan_duty_cycle}, Margelis {\it et al.} \cite{70_low_throughput_networks_IoT} and Danish {\it et al.} \cite{107_blockchain_authentication} stated that join request messages are not encrypted, which means that jamming and replay attacks can occur. D\"onmez and Nigussie added that not only join request messages but also rejoin request messages are not encrypted \cite{02_security_procedure_LoRaWANv1.1}. As a result, a potential eavesdropper can collect the data and run the authentication procedure to integrate the network through a replay attack.

\item Q2 - Na {\it et al.} recommended masking every message with different tokens to prevent a replay attack \cite{44_countermeasure_replay_attack}. This procedure consists of generating a countermeasure with an XOR operation and creating a single join request message that prevents the message from being exploited. Benkahla {\it et al.} suggested using AppKey to encrypt the join request and protect the LoRaWAN information \cite{49_analysis_lorawan_duty_cycle}. Danish {\it et al.} designed a two-factor authentication mechanism to encrypt the join request messages \cite{107_blockchain_authentication}. The framework consists of an independent blockchain network, working concurrently with the LoRaWAN network that runs smart contracts and saves the end device information, which is triggered by the transactions from the gateways.

\item Q3 - Eldefrawy {\it et al.} conducted a LoRaWAN security analysis with a Scyther-proof security protocol verification tool, which takes to account the security of communication messages and their linked keys \cite{01_formal_security_analysis_LoRaWAN}. Na {\it et al.} tested the replay attack in a LoRa end device from a malicious device based on Raspberry Pi 2 Model B and Raspbian-Jessie \cite{44_countermeasure_replay_attack}. Benkahla {\it et al.} carried out a feasibility study to test a proposed solution with a prototype designed with the Libelium Waspmote SX1272-LoRa and a gateway \cite{49_analysis_lorawan_duty_cycle}. Danish {\it et al.} simulated the planned framework using the Ethereum blockchain and Python client through the web and implementation of end device and network servers, respectively \cite{107_blockchain_authentication}. The simulation showed that the required improvement is feasible; however, there was no evaluation of the security.

\item Q4 - LoRaWAN specification versions 1.0 and 1.1.
\end{itemize}

\subsection{Join accept is not linked to join request}

\begin{itemize}
\item Q1 - Eldefrawy {\it et al.} \cite{01_formal_security_analysis_LoRaWAN}, D\"onmez and Nigussie \cite{03_security_LoRaWANv1.1_scenarios} and Van Es {\it et al.} \cite{92_DoS_lorawan} stated that a join accept message is not linked to the related join request message. In this case, when there is a multiple join request, it is not possible to determine which join request is related to the join accept. The end device might lack agreement and synchronization properties with servers, which can lead to impersonating.

\item Q2 - Van Es {\it et al.} pointed out that the problem of vulnerability was solved in specification version 1.1 by including the JoinNonce value \cite{92_DoS_lorawan}.

\item Q3 - Eldefrawy {\it et al.} conducted a formal protocol security analysis with the Scyther-proof security protocol verification tool designed for the security of communication messages and their linked keys \cite{01_formal_security_analysis_LoRaWAN}. Van Es {\it et al.} simulated Coloured Petri Net models with CPNTools to assess the success of a denial-of-service attack \cite{92_DoS_lorawan}. 

\item Q4 - LoRaWAN specification version 1.0.
\end{itemize}

\subsection{AppSKey is derived from AppKey using the network server}

\begin{itemize}
\item Q1 - D\"onmez and Nigussie \cite{02_security_procedure_LoRaWANv1.1}, Naoui {\it et al.} \cite{35_third_party_key_management}, Xia {\it et al.} \cite{56_session_key_management}, Bala {\it et al.} \cite{90_separate_session_key_generation} and Naoui {\it et al.} \cite{98_lorawan_framework_smarthome} stated that the network server derives the application session key from AppKey and thus enables the network server to decrypt data. However, if the network server is not a trusted entity, then there is a risk of data confidentiality and integrity vulnerability.

\item Q2 - D\"onmez and Nigussie pointed out that in specification version 1.1, the join procedure is delegated by introducing a special case for join procedure and its delegation \cite{02_security_procedure_LoRaWANv1.1}. Naoui {\it et al.} \cite{35_third_party_key_management} and Bala {\it et al.} \cite{90_separate_session_key_generation} put forward 2 different root keys: one for end device and network server communication and the other one for end device and application server communication to prevent attacks from a malicious network server. Naoui {\it et al.} \cite{35_third_party_key_management} and Xia {\it et al.} \cite{56_session_key_management} suggested adding a trusted third entity in the LoRaWAN architecture that would be responsible for key management service. Naoui {\it et al.} insisted that the application server must be a trusted entity and be able to generate the NwkSKey and send it to the network server \cite{98_lorawan_framework_smarthome}.

\item Q3 - Naoui {\it et al.} conducted a formal analysis of the LoRaWAN protocol using C/Java-like syntax in the Scyther tool and comparing version 1.0 with the improved solution based on security properties and key management requirements \cite{35_third_party_key_management}. Bala {\it et al.} carried out a check of secrecy and provided authentication of the proposed protocol using the ProVerif automated validation tool \cite{90_separate_session_key_generation} .

\item Q4 - LoRaWAN specification version 1.0.
\end{itemize}

\subsection{LoRaWAN does not feature over-the-air device firmware update}

\begin{itemize}
\item Q1 - D\"onmez and Nigussie stated that LoRaWAN does not have the over-the-air device firmware updates \cite{02_security_procedure_LoRaWANv1.1}, which means that it is expensive to update firmware and also that different versions of the protocol will coexist in a network. In addition, the backward compatibility scenario gives rise to new vulnerabilities.

\item Q2 - No improvement is suggested for this topic.

\item Q3 - No verification or tests were found for this topic.

\item Q4 - LoRaWAN specification versions 1.0 and 1.1.
\end{itemize}

\subsection{The key update is not specified}

\begin{itemize}
\item Q1 - Eldefrawy {\it et al.} \cite{01_formal_security_analysis_LoRaWAN}, Sanchez-Iborra {\it et al.} \cite{31_internet_access_lorawan_devices}, Xia {\it et al.} \cite{56_session_key_management}, and Naoui {\it et al.} \cite{57_session_key_management} argued that it is necessary to update the key periodically to ensure high security resilience, although LoRaWAN does not specify the interval needed for the update. Naoui {\it et al.} \cite{35_third_party_key_management}, Yang {\it et al.} \cite{37_security_vulnerabilities_lorawan}, Skorpil {\it et al.} \cite{59_IoT_security_overview_practical_demonstration} Gresak and Voznak \cite{97_protecting_gateway}, and Naoui {\it et al.} \cite{98_lorawan_framework_smarthome} found that the activation of an end device by personalization does not update the key storaged and after resetting the end device or counter overflow, the frame counter value starts from zero again with the same keys. D\"onmez and Nigussie \cite{109_key_management_healthcare} noted that LoRaWAN has lifetime root keys in the over-the-air activation procedure which exceeds the recommended number of cryptoperiods.

\item Q2 - Sanchez-Iborra {\it et al.} recommended implementing an EDHOC IPv6-based function within the LoRaWAN architecture to enable the inclusion of a periodic key update \cite{31_internet_access_lorawan_devices}. Naoui {\it et al.} proposed updating the AppKey in each session \cite{35_third_party_key_management}, while Yang suggested updating the key whenever the counter has an overflow {\it et al.} \cite{37_security_vulnerabilities_lorawan}. Xia {\it et al.} stated that the session key updates occur automatically and remotely in a key management entity \cite{56_session_key_management}. Naoui {\it et al.} enhanced the LoRaWAN protocol by replacing the 128-bit AES key with a session key \cite{57_session_key_management}. This key will be established between the end device and the provided network server that the protocol allows for two Internet of Things devices in a local network (that are subject to energy and computing constraints) to create a common session key with the help of a subset of neighbors. Gresak and Voznak recommended creating an algorithm in the gateway that could determine if the last message received is the same as the message after the end device restarts \cite{97_protecting_gateway}. Naoui {\it et al.} proposed updating keys after resetting the end device through a one-time password generator \cite{98_lorawan_framework_smarthome}. D\"onmez and Nigussie designed an assisted mode for key management, which allows for the root key updates by introducing a master device in the network \cite{109_key_management_healthcare}.

\item Q3 – Naoui {\it et al.} conducted a formal analysis of the LoRaWAN protocol using C/Java-like syntax in the Scyther tool and comparing version 1.0 with the improved solution based on security properties and key management requirements \cite{35_third_party_key_management}. Yang {\it et al.} experimented with a LoRaWAN provider, which involved operating a malicious gateway based on the popular RN2483 chip to perform the attack \cite{37_security_vulnerabilities_lorawan}. Gresak and Voznak conducted a test for the attack, which included the implemented algorithm \cite{97_protecting_gateway}. The tests were carried out in a real infrastructure currently located in the Czech Republic. Naoui {\it et al.} validated the proposal with a formal verification using the Scyther tool \cite{98_lorawan_framework_smarthome}.

\item Q4 - LoRaWAN specification versions 1.0 and 1.1.

\end{itemize}

\subsection{Counter mode encryption executes only XOR operation}

\begin{itemize}
\item Q1 - Yang {\it et al.} \cite{37_security_vulnerabilities_lorawan}, Lee {\it et al.} \cite{40_countermeasure_bitflipping} and Benkahla {\it et al.} \cite{49_analysis_lorawan_duty_cycle} stated that counter mode only carries out the XOR operation to encrypt plaintext, which retains the same bit order as ciphertext and thus makes it possible to launch a bit-flipping or eavesdropping attack.

\item Q2 - Yang {\it et al.} suggested using different counters to replace the counter value for a valid nonce \cite{37_security_vulnerabilities_lorawan}. Lee {\it et al.} employed a shuffling method of two phases with a shift phase and swap phase to avoid bit-flipping \cite{40_countermeasure_bitflipping}. Benkahla {\it et al.}  suggested shuffling the location of all the bytes in the message and a hash function to add a secret key using the AES offset codebook mode \cite{49_analysis_lorawan_duty_cycle}.

\item Q3 - Lee {\it et al.} conducted experiments with Go language to validate the recommended shuffling method \cite{40_countermeasure_bitflipping}. Benkahla {\it et al.}  demonstrated the feasibility of finding a solution through a prototype that combined the Libelium Waspmote SX1272-LoRa with a gateway \cite{49_analysis_lorawan_duty_cycle}.

\item Q4 - LoRaWAN specification version 1.0.
\end{itemize}

\subsection{The message integrity code has a short length}

\begin{itemize}
\item Q1 - Lee {\it et al.} \cite{40_countermeasure_bitflipping}, Iqbal and Iqbal \cite{64_AES_cryptography} and Raad {\it et al.} \cite{104_elliptic_curve_crypto} showed that cipher-based message authentication code mode generates a 16-bytes result, but since only 4 bytes are included in the message integrity code, the frame payload might be exposed to a brute-force and bit-flipping attack. Coman {\it et al.} pointed out that a malicious entity can send forged messages to the network server and force it to forward a message to the application server \cite{110_IoT_vulnerability_analysis}.

\item Q2 - Raad {\it et al.} suggested adding one more cryptographic layer with a digital signature by employing elliptic curve methods \cite{104_elliptic_curve_crypto}.

\item Q3 - Lee {\it et al.} carried out a message integrity code simulation in Go language that makes the time measurement recalculate the message integrity code after a change in ciphertext \cite{40_countermeasure_bitflipping}. They showed that each change takes $9287.2$ ms. Iqbal and Iqbal conducted a benchmark test of a brute-force attack on an AES using off-the-shelf components that proved the AES is vulnerable \cite{64_AES_cryptography}. Raad {\it et al.} checked the efficiency of a developed adaptive method and compared it with the Koblitz method by taking note of the time response for encryption and decryption \cite{104_elliptic_curve_crypto}. The tests were conducted using a Raspberry Pi3B as a network server installed in a Linux system, with an RHF0m301 as a gateway. Coman {\it et al.} carried out a test with a Kerlink Wirnet iFemtocell gateway and Microchip RN2483 PICtail Daughter Board applying LoRaWAN 1.0.X that were used to forge messages to the network server hosted in The Things Network \cite{110_IoT_vulnerability_analysis}.

\item Q4 - LoRaWAN specification version 1.0.
\end{itemize}

\subsection{DevNonce management is not specified in detail}

\begin{itemize}
\item Q1 - D\"onmez and Nigussie showed that the network server keeps track of the most recent DevNonce values \cite{02_security_procedure_LoRaWANv1.1}. If a malicious network server replays the same join request message to the join server, then it can saturate the JoinNonce counter and reset the session key. D\"onmez and Nigussie stated that this could occur when the end device is reset or when the counter overflows as a result of activation by personalization, while in over-the-air activation, it is due to counter overflow within a session \cite{03_security_LoRaWANv1.1_scenarios}. Naoui {\it et al.} stated that the management strategy adopted by the network server to check the DevNonce could affect the end device \cite{35_third_party_key_management}. This can occur when the network server just stores some DevNonce, and hence, a previous DevNonce that had already been excluded can be used in a replay attack.
Eldefrawy {\it et al.} \cite{01_formal_security_analysis_LoRaWAN} and Margelis {\it et al.} \cite{70_low_throughput_networks_IoT} pointed out that DevNonce is not uniformly random. Kim and Song \cite{21_analysis_LoRaWAN_v1.1_security}, Tomasin {\it et al.} \cite{34_lorawan_join_procedure}, Sung {\it et al.} \cite{45_protecting_end_device} and Xia {\it et al.} \cite{56_session_key_management} added that the same DevNonce can be created more than once by the same end device, depending on the way the network server deals with it, and thus, the genuine end device can be excluded from the network and suffer a replay attack.

\item Q2 - D\"onmez and Nigussie put forward countermeasures that can enable the join server to keep track of DevNonce \cite{02_security_procedure_LoRaWANv1.1}. Kim and Song \cite{21_analysis_LoRaWAN_v1.1_security} set out a strategy that has a different join request scheme: whether NwkSKey is available or not. In a noninitial join request, NwkSKey is used to calculate the message integrity code instead of AppKey as NwSKey changes in each join procedure and AppKey is the same throughout the life of the end device life, and the new scheme strengthens the security resilience of LoRaWAN. Moreover, the duplicated DevNonce is less likely to occur, as it is only taken into account if there is an initial join request, which is rarely the case. Tomasin {\it et al.} argued that DevNonce is a sequential number that is used to avoid random uniformity in the join procedure \cite{34_lorawan_join_procedure}. Naoui {\it et al.} \cite{35_third_party_key_management} and Xia {\it et al.} \cite{56_session_key_management} suggested replacing the DevNonce with timestamp, which means that only the last timestamp had to be stored.

\item Q3 - Kim and Song confirmed they could improve the DevNonce duplication by adopting a mathematical approach \cite{21_analysis_LoRaWAN_v1.1_security}. This idea was also supported by using an SK-IM880B board with 16-bit DevNonce and employing the Semtech algorithm to create DevNonce. Tomasin {\it et al.} demonstrated that by adopting a mathematical approach, there is a high probability that DevNonce can be duplicated \cite{34_lorawan_join_procedure}. Naoui {\it et al.} conducted a formal analysis of how the LoRaWAN protocol can be established using C/Java-like syntax in the Scyther tool and by comparing version 1.0 and the improved solution that is based on security properties and critical management requirements \cite{35_third_party_key_management}.

\item Q4 - LoRaWAN specification version 1.0.
\end{itemize}

\subsection{The management of frame counters is not specified in detail}

\begin{itemize}
\item Q1 - D\"onmez and Nigussie added that replay attack could occur in case the end device is reset or the counter overflows \cite{03_security_LoRaWANv1.1_scenarios}. Aras {\it et al.} stated that if the frame counters are not managed correctly, then the message can be resent in a replay attack \cite{32_security_vulnerabilities_LoRa}. Skorpil {\it et al.} \cite{59_IoT_security_overview_practical_demonstration} and Kail {\it et al.} \cite{77_security_survey} argued that the frame counters in application layers do reset if the server needs a reboot, and in that case, it is possible to launch a replay attack.

\item Q2 - Aras {\it et al.} put forward a solution that involves application developers not carrying out sensitive operations \cite{32_security_vulnerabilities_LoRa}.

\item Q3 - Skorpil {\it et al.} showed that a replay attack activated by personalization node is successfully conducted \cite{59_IoT_security_overview_practical_demonstration}. The first part of the Internet of Things system is a SolidusTECH mini UNI sensor activated by personalization. The tests proceeded by building a whole Internet of Things system with off-the-shelf components. In addition, Python scripts were used to sniff the communication on the server and replicate the traffic between the sensor and gateway to launch the attack.

\item Q4 - LoRaWAN specification versions 1.0 and 1.1.
\end{itemize}

\subsection{End devices are not linked to a particular gateway}

\begin{itemize}
\item Q1 - Hellebrandt {\it et al.} claimed that it is possible to set up a malicious entity near the end device so that the messages could be received as a gateway, given the fact that end devices are not linked to a gateway \cite{16_privacy_ioT_networks}.

\item Q2 - Hellebrandt {\it et al.} recommended building a private network and installing an anonymization IP address \cite{16_privacy_ioT_networks}.

\item Q3 - No verification or tests were found for this topic.

\item Q4 - LoRaWAN specification version 1.0.
\end{itemize}

\subsection{A downlink message does not have a defined route}

\begin{itemize}
\item Q1 - Gemein \cite{23_lora_secure_design} and Van Es {\it et al.} \cite{92_DoS_lorawan} argued that when the network server has to send the downlink from the end device and it has multiple gateway alternatives, it must choose one gateway which makes it possible to handle the routing.

\item Q2 - Tomar and Gemein also suggested including additional keys to allow for messages to be routed actively \cite{23_lora_secure_design}. Van Es {\it et al.} proposed implementing handshaking between the network server and end device to ensure the successful delivery of downlink messages \cite{92_DoS_lorawan}.

\item Q3 - Van Es {\it et al.} simulated Coloured Petri Net models of downlink routing attacks with CPNTools to determine the success of a denial-of-service attack \cite{92_DoS_lorawan}.

\item Q4 - LoRaWAN specification version 1.0.
\end{itemize}

\subsection{Beacons are not encrypted}

\begin{itemize}
\item Q1 - Yang {\it et al.} \cite{37_security_vulnerabilities_lorawan} and Van Es {\it et al.} \cite{92_DoS_lorawan} revealed that since beacons for class B end devices are not encrypted, an attacker can send out a beacon with malicious parameters to the end devices to desynchronize it.

\item Q2 - Yang {\it et al.} also suggested adding a physical layer cyclic redundancy check in the message integrity code of cryptographic signatures \cite{37_security_vulnerabilities_lorawan}. Van Es {\it et al.} argued that beacons should be authenticated to avoid denial-of-service, but they did not specify an authentication method \cite{92_DoS_lorawan}.

\item Q3 - Van Es {\it et al.} simulated Coloured Petri Net models of attack with CPNTools to determine the success of a denial-of-service attack \cite{92_DoS_lorawan}.

\item Q4 - LoRaWAN specification versions 1.0 and 1.1.
\end{itemize}

\subsection{Receive windows are not controlled}

\begin{itemize}
\item Q1 - Mikhaylov {\it et al.} stated that receive windows are mandatory in LoRaWAN end devices, although the end device does not control them; moreover, the way that the design has to follow the specification is easily predicted \cite{99_energy_attack}. In this way, an attacker can send a message to the end devices in the two receive windows that are opened after transmission. This will increase the power consumption as the packets are checked just after the entire message has been received, which can allow for an energy attack to occur.

\item Q2 - No improvements are suggested for this topic.

\item Q3 - Mikhaylov {\it et al.} tested the energy attack on a LoRaWAN end device prototype, and it found that end device power consumption can increase from $36$\% to $576$\% \cite{99_energy_attack}.

\item Q4 - LoRaWAN specification version 1.1.
\end{itemize}

\subsection{Acknowledge data messages are not linked to the related data message}

\begin{itemize}
\item Q1 - D\"onmez and Nigussie \cite{03_security_LoRaWANv1.1_scenarios}, Yang {\it et al.} \cite{37_security_vulnerabilities_lorawan} and Skorpil {\it et al.} \cite{59_IoT_security_overview_practical_demonstration} stated that there is no mechanism that can combine the received acknowledgment messages with the acknowledge data messages. Skorpil {\it et al.} also stated that a malicious entity could keep possession of a message and use it later in a replay attack \cite{59_IoT_security_overview_practical_demonstration}.

\item Q2 - Yang {\it et al.} proposed adding a cryptographic checksum with the returned acknowledgment or authenticated encryption of the data link layer payload \cite{37_security_vulnerabilities_lorawan}.

\item Q3 - Yang {\it et al.} performed a selective acknowledge spoofing attack by assuming that the gateway is malicious and selectively suppressing specific frames from transmission \cite{37_security_vulnerabilities_lorawan}.

\item Q4 - LoRaWAN specification version 1.0.
\end{itemize}

\subsection{Downlink messages do not have checksum over the payload}

\begin{itemize}
\item Q1 - Margelis {\it et al.} stated that download messages do not have checksum, which can lead to integrity attacks \cite{70_low_throughput_networks_IoT}.

\item Q2 - No improvement is suggested for this topic.

\item Q3 - No verification or tests were found for this topic.

\item Q4 - LoRaWAN specification version 1.0.
\end{itemize}

\subsection{AppNonce is not registered by end device}

\begin{itemize}
\item Q1 - D\"onmez and Nigussie \cite{03_security_LoRaWANv1.1_scenarios}, Naoui {\it et al.} \cite{35_third_party_key_management}, and Naoui {\it et al.} \cite{98_lorawan_framework_smarthome} stated that AppNonce is not registered by the end device for join accept messages. This means that an attacker can reuse the AppNonce to desynchronize end device and the network server with different session keys.

\item Q2 - Naoui {\it et al.} proposed exchanging nonce with timestamps \cite{98_lorawan_framework_smarthome}.

\item Q3 - No verification or tests were found for this topic.

\item Q4 - LoRaWAN specification version 1.0.
\end{itemize}

\subsection{The join procedure does not specify the number of unconfirmed sessions that are needed for a network server storage per end device}

\begin{itemize}
\item Q1 - D\"onmez and Nigussie stated that join procedure does not specify the number of unconfirmed sessions required for a network server storage per end device \cite{03_security_LoRaWANv1.1_scenarios}. If a network server rejects the former context without waiting for a confirmation from the end device, then the end device and the network server end up using different security contexts. 

\item Q2 - No improvement is suggested for this topic.

\item Q3 - No verification or tests were found for this topic.

\item Q4 - LoRaWAN specification version 1.0.

\end{itemize}

\subsection{Key provisioning, storage and usage is not specified}

\begin{itemize}
\item Q1 - Eldefrawy {\it et al.} stated that the preloading root keys violate the key-control property \cite{01_formal_security_analysis_LoRaWAN}. D\"onmez and Nigussie stated that root key provisioning, storage and usage are beyond the scope of the specifications and pose a serious challenge \cite{02_security_procedure_LoRaWANv1.1}. Gladisch {\it et al.} pointed out that since there are no specifications for key provisions, storage and use, the implementation must provide the means for guaranteeing security \cite{24_securely_IoT_lorawan}. Margelis {\it et al.} stated that the generation method of activation by personalization keys is not specified \cite{70_low_throughput_networks_IoT}.

\item Q2 - No improvement is suggested for this topic.

\item Q3 - No verification or tests were found for this topic.

\item Q4 - LoRaWAN specification versions 1.0 and 1.1.
\end{itemize}

\subsection{Discussion}

This systematic review has drawn on thirty-one primary studies to answer the defined question; $77.5$\% of these studies are related to LoRaWAN 1.0 version, and $22.5$\% are associated with LoRaWAN 1.1 version. All the primary studies assisted in detecting vulnerabilities, $71$\% of the primary studies made suggestions for improvement, and $58.1$\% tested or checked a security vulnerability and what improvement could be made. Table \ref{systematic_review_results} shows all the results of the systematic review.

\begin{table}[!ht]
\centering
\caption{Systematic Review Results}
\label{systematic_review_results}
    \begin{tabular}{p{0.5\textwidth}|p{0.08\textwidth}|p{0.08\textwidth}|p{0.07\textwidth}|p{0.03\textwidth}}
    \hline
    \textbf{Item}  & \textbf{Q1} & \textbf{Q2}  & \textbf{Q3} & \textbf{Q4} \\ \hline 
    
     A lack of end-to-end integrity protection 
     & \cite{01_formal_security_analysis_LoRaWAN} 
     \cite{03_security_LoRaWANv1.1_scenarios}
     \cite{24_securely_IoT_lorawan} \cite{37_security_vulnerabilities_lorawan}
     \cite{59_IoT_security_overview_practical_demonstration} \cite{98_lorawan_framework_smarthome}
     & \cite{24_securely_IoT_lorawan} \cite{37_security_vulnerabilities_lorawan}
     \cite{98_lorawan_framework_smarthome}
     & \cite{37_security_vulnerabilities_lorawan}
     & 1.0 1.1 \\ \hline
     
    Join and rejoin request messages are not encrypted 
    & \cite{01_formal_security_analysis_LoRaWAN}   \cite{35_third_party_key_management} \cite{44_countermeasure_replay_attack}    \cite{49_analysis_lorawan_duty_cycle} \cite{70_low_throughput_networks_IoT} \cite{107_blockchain_authentication} 
    & \cite{44_countermeasure_replay_attack} \cite{49_analysis_lorawan_duty_cycle} \cite{107_blockchain_authentication} 
    &  \cite{01_formal_security_analysis_LoRaWAN} \cite{44_countermeasure_replay_attack} \cite{107_blockchain_authentication} 
    & 1.0 1.1 \\ \hline

    Join accept is not linked to join request 
    &  \cite{01_formal_security_analysis_LoRaWAN} \cite{03_security_LoRaWANv1.1_scenarios}    \cite{92_DoS_lorawan} 
    & \cite{92_DoS_lorawan} 
    &  \cite{01_formal_security_analysis_LoRaWAN}    \cite{92_DoS_lorawan} 
    & 1.0 \\ \hline 

    AppSKey is derived from AppKey using the network server
    &  \cite{02_security_procedure_LoRaWANv1.1} \cite{35_third_party_key_management} 
    \cite{56_session_key_management} \cite{90_separate_session_key_generation} \cite{98_lorawan_framework_smarthome} 
    & \cite{02_security_procedure_LoRaWANv1.1}  \cite{35_third_party_key_management}  \cite{56_session_key_management} \cite{90_separate_session_key_generation}  \cite{98_lorawan_framework_smarthome}  
    & \cite{35_third_party_key_management}   \cite{90_separate_session_key_generation}  
    & 1.0 \\ \hline 
   
    LoRaWAN does not feature over-the-air device firmware update 
    & \cite{02_security_procedure_LoRaWANv1.1} 
    & 
    & 
    & 1.0 1.1 \\ \hline  

    The key update is not specified
    & \cite{01_formal_security_analysis_LoRaWAN}
    \cite{31_internet_access_lorawan_devices}
    \cite{35_third_party_key_management}
    \cite{37_security_vulnerabilities_lorawan}
    \cite{56_session_key_management}
    \cite{57_session_key_management}
    \cite{59_IoT_security_overview_practical_demonstration}
    \cite{97_protecting_gateway}
    \cite{98_lorawan_framework_smarthome}
    \cite{109_key_management_healthcare}
    &  \cite{31_internet_access_lorawan_devices}
    \cite{35_third_party_key_management}
    \cite{37_security_vulnerabilities_lorawan}
    \cite{56_session_key_management}
    \cite{57_session_key_management}
    \cite{97_protecting_gateway}
    \cite{98_lorawan_framework_smarthome}
    \cite{109_key_management_healthcare}
    & \cite{35_third_party_key_management}
    \cite{37_security_vulnerabilities_lorawan}
    \cite{97_protecting_gateway}
    \cite{98_lorawan_framework_smarthome}
    & 1.0 1.1  \\ \hline  
    
    Counter mode encryption only carries out the XOR operation & \cite{37_security_vulnerabilities_lorawan} \cite{40_countermeasure_bitflipping}   \cite{49_analysis_lorawan_duty_cycle} 
    & \cite{37_security_vulnerabilities_lorawan}   \cite{40_countermeasure_bitflipping} \cite{49_analysis_lorawan_duty_cycle} 
    &  \cite{40_countermeasure_bitflipping} \cite{49_analysis_lorawan_duty_cycle} 
    & 1.0 \\ \hline
    
    The message integrity code has a short length
    & \cite{40_countermeasure_bitflipping}  
    \cite{64_AES_cryptography} 
    \cite{104_elliptic_curve_crypto}
    \cite{110_IoT_vulnerability_analysis}
    & \cite{104_elliptic_curve_crypto}
    &  \cite{40_countermeasure_bitflipping}  
    \cite{64_AES_cryptography} 
    \cite{104_elliptic_curve_crypto}
    \cite{110_IoT_vulnerability_analysis}
    & 1.0 \\ \hline
    
    DevNonce management is not specified in detail
    & \cite{01_formal_security_analysis_LoRaWAN} 
    \cite{02_security_procedure_LoRaWANv1.1} \cite{03_security_LoRaWANv1.1_scenarios} 
    \cite{21_analysis_LoRaWAN_v1.1_security}
    \cite{24_securely_IoT_lorawan}
    \cite{34_lorawan_join_procedure}
    \cite{35_third_party_key_management} 
    \cite{56_session_key_management}
    \cite{70_low_throughput_networks_IoT}
    &  \cite{02_security_procedure_LoRaWANv1.1}
    \cite{21_analysis_LoRaWAN_v1.1_security}
    \cite{34_lorawan_join_procedure}
    \cite{35_third_party_key_management} 
    \cite{56_session_key_management}
    & \cite{21_analysis_LoRaWAN_v1.1_security}
    \cite{34_lorawan_join_procedure} \cite{35_third_party_key_management}
    & 1.0  \\ \hline

    The management of frame counters is not specified in detail
    & \cite{03_security_LoRaWANv1.1_scenarios} \cite{32_security_vulnerabilities_LoRa}   \cite{59_IoT_security_overview_practical_demonstration} 
    \cite{77_security_survey}
    & \cite{32_security_vulnerabilities_LoRa}  
    & \cite{59_IoT_security_overview_practical_demonstration} 
    & 1.0 1.1  \\ \hline 

    End devices are not linked to a particular gateway
    & \cite{16_privacy_ioT_networks} 
    & \cite{16_privacy_ioT_networks} 
    &  
    & 1.0 \\ \hline
    
    A downlink message does not have a defined route
    & \cite{23_lora_secure_design}
    \cite{92_DoS_lorawan}
    & \cite{23_lora_secure_design}
    \cite{92_DoS_lorawan}
    &  \cite{92_DoS_lorawan}
    & 1.0 \\ \hline
    
    Beacons are not encrypted  
    & \cite{37_security_vulnerabilities_lorawan} 
    \cite{92_DoS_lorawan}   
    & \cite{37_security_vulnerabilities_lorawan}  
    \cite{92_DoS_lorawan} 
    & \cite{92_DoS_lorawan} 
    & 1.0 1.1 \\ \hline
    
    Receive windows are not controlled 
    & \cite{99_energy_attack} 
    &
    & \cite{99_energy_attack} 
    & 1.1 \\ \hline
    
    Acknowledge data messages are not linked to the related data message  
    & \cite{03_security_LoRaWANv1.1_scenarios} \cite{37_security_vulnerabilities_lorawan} \cite{59_IoT_security_overview_practical_demonstration} 
    & \cite{37_security_vulnerabilities_lorawan}
    & \cite{37_security_vulnerabilities_lorawan}
    & 1.0 \\ \hline
    
    Downlink messages do not have checksum over the payload  
    & \cite{70_low_throughput_networks_IoT} 
    &         
    &  
    & 1.0 \\ \hline
    
    AppNonce is not registered by end device 
    & \cite{03_security_LoRaWANv1.1_scenarios}  \cite{35_third_party_key_management} \cite{98_lorawan_framework_smarthome} 
    & \cite{98_lorawan_framework_smarthome} 
    & 
    & 1.0 \\ \hline
    
    The join procedure does not specify the number of unconfirmed sessions that are needed for a network server storage per end device 
    & \cite{03_security_LoRaWANv1.1_scenarios} 
    &  
    & 
    & 1.0 \\ \hline

    Key provisioning, storage, and usage is not specified
    & \cite{01_formal_security_analysis_LoRaWAN}
    \cite{02_security_procedure_LoRaWANv1.1}
    \cite{24_securely_IoT_lorawan}
    \cite{70_low_throughput_networks_IoT}
    &  
    & 
    & 1.0 1.1 \\ \hline
    
    \end{tabular}
\end{table}

The LoRaWAN protocol is based on the LoRa physical layer that is wireless technology. Hence, the inherent vulnerabilities of wireless communication are all included in LoRaWAN applications (such as jamming and traffic analysis). The LoRaWAN specification consists of three types of end devices: Class A, Class B, and Class C. 
Almost the whole of the security analysis is concerned with basic Class A end device requirements; there are also a few data regarding the specific needs of Class B, but no study was found for the Class C end device. One crucial point is that the LoRaWAN protocol does not support firmware updates, and an application will consist of components of versions 1.0 and 1.1. Thus the compatibility scenario must be taken into account in the security analysis. The LoRaWan specification provides activation by personalization and over-the-air activation as an authentication process, and most of the discovered vulnerabilities are related to the join procedure. About the systematic review, it should also be noted that the checking or testing that is carried out tend to focus on vulnerability and to make improvements to the solution. Some vulnerabilities that were found in version 1.0 were addressed in version 1.1, and for this reason, specification version 1.1 provides better security resilience than does version 1.0. However, the in-depth security analysis is needed. Moreover, the implemented modifications can generate new types of vulnerability that will mainly apply to version 1.1.

\section{Conclusions}
This paper presented a systematic review according to Kitchenham methodology related to security in the LoRaWAN protocol \cite{kitchenham}. In the planning phase, four questions were designed, the string search \textit{(Security OR Cybersecurity) AND (LoRa OR LoRaWAN) AND Network} was selected, and the review protocol defined. In the conducting phase, the search was carried out and returned one hundred and twenty-three studies of digital libraries from Springer Link, Science Direct, IEEE Xplore, and ACM. Additionally, in the conducting phase, thirty-nine primary studies are selected; their data were extracted according to outlined questions. The quality assessment of the primary studies showed the high quality of the extracted data. Finally, in the reporting phase, the extracted data were compiled and organized according to related vulnerability.

However, it was possible to detect vulnerabilities within the LoRaWAN specification that are related to the protocol specification and implementation techniques, as well as cryptographic algorithms. As LoRaWAN is a relatively new protocol, its level of security has not yet been rigorously studied in the literature. Future studies should focus on LoRaWAN specification version 1.1, the compatibility scenario for end device version 1.0 and 1.1, Class B and C end devices, the implementation of the protocol, and bench testing.

\begin{acks}
This research is part of the INCT of the Future Internet for Smart Cities funded by CNPq proc. 465446/2014-0, Coordenação de Aperfeiçoamento de Pessoal de Nível Superior – Brasil (CAPES) – Finance Code 001, FAPESP proc. 14/50937-1, and FAPESP proc. 15/24485-9.
\end{acks}

\bibliographystyle{plain}
\bibliography{acmart}

\end{document}